\documentclass[a4paper,fleqn,usenatbib]{mnras}

\usepackage{newtxtext,newtxmath}
\usepackage[T1]{fontenc}
\usepackage{ae,aecompl}

\usepackage{graphicx}	% Including figure files
\usepackage{amsmath}	% Advanced maths commands
\usepackage{amssymb}	% Extra maths symbols

\title[Structural diversity of disc galaxies]{Structural diversity of disc galaxies originating in the cold gas inflow from cosmic webs}

\author[M. Noguchi]{
Masafumi Noguchi $^{1}$\thanks{E-mail: noguchi@astr.tohoku.ac.jp (MN)}
\\
$^{1}$Astronomical Institute, Tohoku University, 6-3, Aramaki, Aoba-ku, Sendai, Miyagi, 980-8578, Japan
}

\date{Accepted XXX. Received YYY; in original form ZZZ}

\pubyear{2019}

\begin{document}
\label{firstpage}
\pagerange{\pageref{firstpage}--\pageref{lastpage}}
\maketitle

\begin{abstract}
	Disc galaxies show a large morphological diversity with varying contribution of 
    three major structural components; thin discs, thick discs, and central bulges. Dominance of bulges increases
	with the galaxy mass (Hubble sequence) whereas thick discs are more prominent in lower mass galaxies.
	Because galaxies grow with the accretion of matter, this observed variety should reflect diversity
	in accretion history. On the basis of the prediction by the cold-flow theory for galactic gas accretion
	and inspired by the results of previous studies,
	we put a hypothesis that associates different accretion modes with different components. 
	Namely, thin discs form as the shock-heated hot gas in high-mass halos
    gradually accretes to the central part,
	thick discs grow by the direct accretion of cold gas from 
	cosmic webs when the halo mass is low, and finally bulges form by the inflow of cold gas
	through the shock-heated gas in high-redshift massive halos.
	We show that this simple hypothesis reproduces the mean observed variation of galaxy morphology with the galaxy mass.
	This scenario also predicts that thick discs are older and poorer in metals than thin discs, 
	in agreement with the currently available observational data.
\end{abstract}


%%%%%%%%%%%%%%%%%%%%%%%%%%%%%%%%%%%%%%%%%%%%%%%%%%

%%%%%%%%%%%%%%%%% BODY OF PAPER %%%%%%%%%%%%%%%%%%

\section{introduction}

Complex structures of disc galaxies pose a tough challenge for galaxy formation theory.
Existence of distinct disc and bulge components and their relative dominance constitutes the basis for the traditional 
Hubble classification of galaxy morphology. Discs are further divided into thin and thick discs with different scale heights \citep[e.g.][]{bu79,ts80}.
Recent studies show that
the low-mass galaxies generally contain significant thick discs but their contribution decreases for higher mass
galaxies \citep[e.g.][]{co14}.
Various properties such as
colours, ages, and metallicities, as well as kinematical properties, are systematically correlated with these respective 
structural components \citep[e.g.][]{mo10}. Despite a large accumulation of past observational and 
theoretical studies \citep[e.g.][]{br14,hu15,br09},
there is not yet a
consensus on how various components of disc galaxies were formed throughout the history of the Universe, and when and how 
the observed diversity of disc galaxy morphology comes out.

Galaxy formation process is greatly controlled by how the gas flows into the dark matter halos and accretes finally onto their central part where stars form.
The longstanding paradigm, the shock-heating theory \citep[e.g.][]{re77}, argues that the gas that entered the halo is heated by a shock wave to a high (virial) temperature, 
and then gradually accretes to the central part as it cools by emitting radiation (cooling flow).
This picture predicts a smooth and continuous gas accretion history. 
Therefore it is not straightforward to understand why and how the different components of disc galaxies formed, and external origins are invoked in some cases. 
For example, bulges are sometimes regarded as smaller galaxies swallowed in galaxy mergers \citep[e.g.][]{za12}. 

The shock-heating theory was later modified by the cold accretion scenario which claims that the shock waves do not arise 
in some cases and a significant part of the gas stays cold, reaching the galaxy in narrow streams almost 
in a freefall fashion (cold accretion) \citep[e.g.][]{fa01,ke05,de06}. This idea was used to interpret a high incidence of actively star-forming galaxies observed at high redshifts \citep{de09}. 
It is also invoked to quench star formation in massive galaxies in cooperation with
the feedback from active galactic nuclei in theoretical models \citep{cr06}. The predicted cold gas streams are claimed to be found in the form
of quasar absorption line systems such as Lyman limit systems \citep{fu11} or Lyman $\alpha$ emitting filaments 
around forming galaxies \citep{ma16}.

Using a simple model for the gas accretion into a dark matter halo and subsequent cooling and star formation processes,
\citet{no18b} applied the cold accretion scenario to the Milky Way (MW), and showed that the existence of two chemically 
different stellar groups observed in the MW disc \citep[e.g.][]{ha13,ha15} is naturally explained.
In this picture, the stellar group rich (relatively to Fe) in $\alpha$  elements (O, Mg, Si, etc) was formed by a fast star formation event fuelled by the cold accretion
in early times whereas the $\alpha$-poor stellar group was gradually formed by the cooling flow accretion of the shock-heated gas 
in later epochs. Interestingly, the $\alpha$-rich stars in the solar neighbourhood tend to have larger velocities perpendicular
to the disc plane than $\alpha$-poor stars \citep[e.g.][]{ha13}, meaning that they reach to higher altitudes resulting into a thicker configuration.
Although this correspondence is not always clear-cut \citep{ha13}, the result of \citet{no18b} hints at the association of 
the cold accretion and the cooling flow with the formation of the MW thick disc and thin disc, respectively 
\citep[also see][]{br09}.
In more general perspective, we can envisage that a particular mode of gas accretion is associated with a 
particular component in disc galaxies.

Here we extend this idea and apply this ``accretion-component correspondence'' to 
the disc galaxy evolution model with various masses, 
and examine if the cold accretion theory explains the observed characteristics (mass fractions, age, metallicity) 
of various structural components across galaxy mass.  As described below, we take an idealized approach in which the model galaxy
evolves ``passively'' under the various accretion modes 
ignoring other potentially important physics (e.g., gas recycling by galactic fountains), 
because we hope to know to what extent 
the build-up of various galactic structures is governed by the accretion process.

\section{connecting accretion to components}

\begin{figure}
	\includegraphics[width=1.1\linewidth]{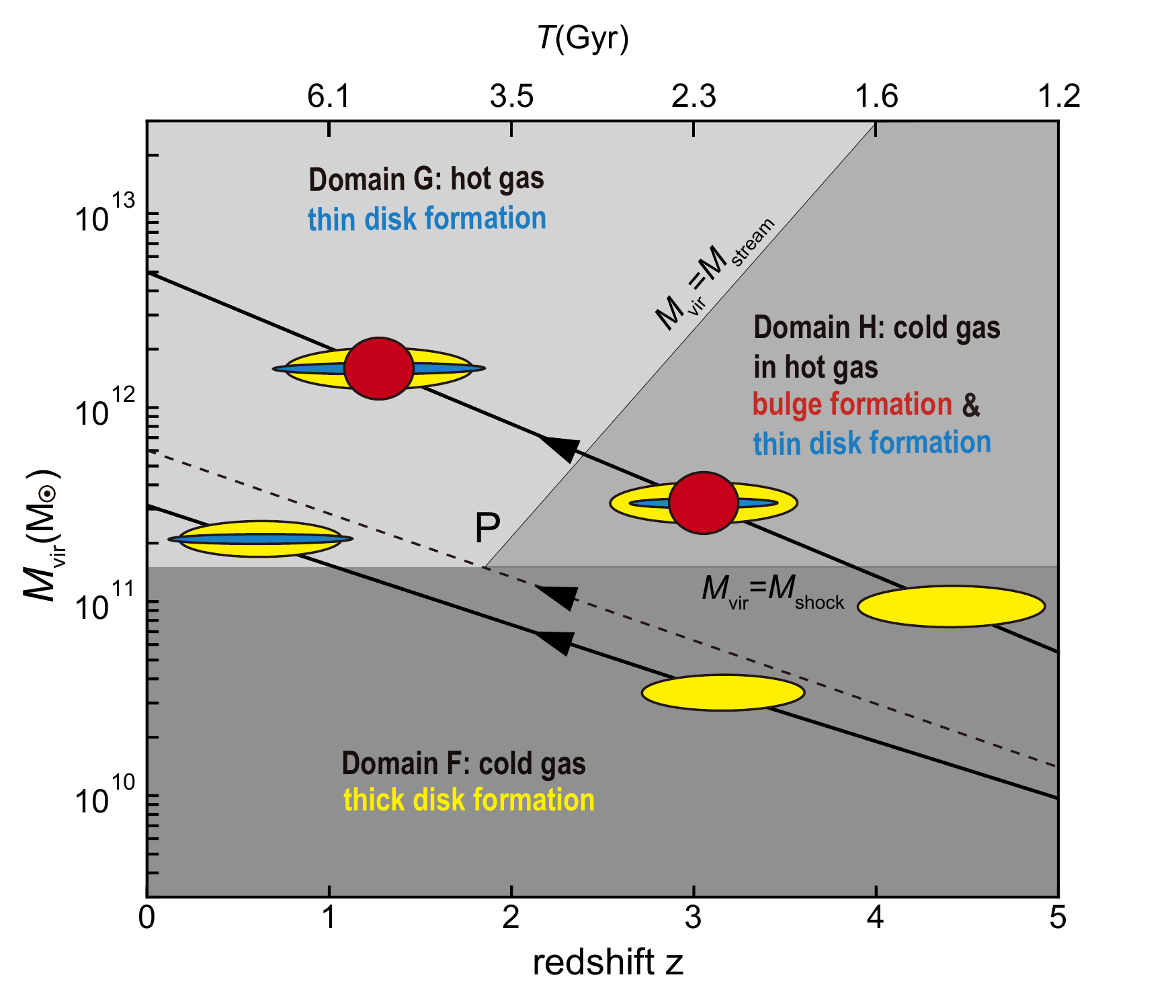}
    \caption{Three different states of the halo gas according to the cold-accretion theory.
	$M_{\rm vir}$, $z$, and $T$ denote the virial mass of the galaxy, redshift, and cosmic time, respectively. 
    $M_{\rm shock}$ and $M_{\rm stream}$ are the mass scales explained in the text.
	Formation of thin disc (blue), thick disc (yellow), and bulge (red) is indicated schematically.
    Two solid diagonal lines indicate the evolutionary path of the least and most massive galaxies studied here.
    The present virial masses for these 
    galaxies are $M_{\rm vir,0}=3.05 \times 10^{11} {\rm M}_{\odot}$ and $4.98 \times 10^{12} {\rm M}_{\odot}$, respectively.
    The dashed line indicates the galaxy which passes through the point P, i.e.,
    the intersection of the $M_{\rm shock}$ line and $M_{\rm stream}$ line. The present virial mass of this galaxy
    (which we call ``the critical mass'') is $M_{\rm critical} = 5.72 \times 10^{11} {\rm M}_{\odot}$. 
    }
    \label{Fig.1}
\end{figure}

The cold-accretion theory predicts three regimes for the thermal state 
and accretion process of the halo gas depending on the virial mass ($M_{\rm vir}$) of the halo and redshift $z$ as
depicted in Fig.1 \citep{de06,oc08}. 
There are two importrant mass scales in defining those regimes. First, $M_{\rm shock}$ is the virial mass
above which a stable shock develops and the halo gas is heated to the virial temperature. 
Second, $M_{\rm stream}$ is the mass above which the narrow streams of cold unshocked gas dissapear at 
high redshifts. Therefore, the cold accretion dominates when $M_{\rm vir}<M_{\rm shock}$ (Domain F), 
whereas the cooling flow dominates in Domain G ($M_{\rm vir}>M_{\rm shock}$  and $M_{\rm vir}>M_{\rm stream}$). 
The halo gas in Domain H ($M_{\rm shock}<M_{\rm vir}<M_{\rm stream}$) has a unique hybrid structure with 
narrow streams of the cold gas penetrating ambient shock-heated gas. We assume that the cold gas and the shocked gas 
experience the cold accretion and the cooling flow, respectively. Here we take $M_{\rm shock} = 1.5 \times 10^{11} {\rm M}_{\odot}$
and ${\log} M_{\rm stream} = 9.2 + 1.067 z$ (see the Supplementary Material for details).

A variant of the evolution model in \citet{no18a} was used in order to examine the growth of various components
 as the halo moves on this plane (see the Supplementary Material for details). In short, the original 
model  was simplified by ignoring the internal structure of the galaxy (``one-zone model'') with each galactic component 
at every moment specified only by its mass.  The gas accretion rate  
to the central galaxy is calculated as a function of time and we apply ``accretion-component correspondence''
in order to trace the mass growth of each galactic component.
As discussed above, we assume that the gas accreted in unheated state (Domain F) ends-up in the thick disc, 
whereas the gas supplied by the cooling flow (Domain G) is used to grow the thin disc. 
The fate of the gas in Domain H is more elusive. By the reason discussed below, we associate this domain with the bulge formation. 
Galactic bulges, including that of MW, are complex structures \citep[e.g.][]{ne13}, and their origin(s) is not yet clear \citep[e.g.][]{ro17}. 
One possibility is that the gravitationally unstable gas-rich discs in young galaxies develop star-forming dense clumps, 
which subsequently sink toward the galactic center due to dynamical friction to form a bulge as first 
proposed by \citet{no98}. 
Alternatively, clumps may form in the cold gas filaments feeding the galaxy \citep{du12}.
This kind of scenarios rely on the longevity of the clumps, which are constantly subject to disruption by feedbacks 
from star formation inside. The hybrid nature of the halo gas in Domain H may help those clumps survive longer 
owing to high external pressure exerted by the surrounding hot gas component. 
Indeed, cosmological simulations for disc galaxy formation suggest the confinement of cold gas stream by
large pressure of the shock-heated gas is a generic phenomenon.
In Figs.4 and 5 of \citet{ke09} and Fig.10 of \citet{ne13}, 
 the streams of cold gas are seen to get thinner 
once they enter the region occupied by the hot shock-heated gas while they reach to the galactic center 
mostly maitaining their widths in the absence of the hot gas.
Also the overdensity of the cold gas is much larger than that of the hot gas in massive galaxies \citep[e.g.][]{va12}, suggesting that
the cold streams are squeezed by the high pressure of the shock-heated gas. 
Clumpy structures are ubiquitous in cold streams. In isolation, they will 
disrupt once a certain amount of stars form in them. However, in the high-pressure environments, they will
survive to reach the central parts of the halos.
Based on this consideration, we assume that the cold gas in Domain H ends up in bulges eventually
whereas the cooling flow of the shock-heated gas in Domain H contributes to thin disc formation. 

By varying the present host halo virial mass $M_{\rm vir,0}$, we examined how the growth of each structural component depends on the galaxy mass.
Nine models are calculated with ${\log}  M_{\rm vir,0}$
    equally spaced between the lowest and highest values indicated in Fig.1.
    Time step of $7 \times 10^6$ years was 
    used to evolve the model, starting at $z=37$. 
    The present total stellar mass of the least and most massive galaxies are 
    $M_{\rm star}=2.13 \times 10^{9} {\rm M}_{\odot}$ and $1.50 \times 10^{11} {\rm M}_{\odot}$, respectively.
    The ``critical'' model is defined as the model which passes through the intersection of 
    the $M_{\rm shock}$ line and $M_{\rm stream}$ line (the dashed line in Fig.1), and its present stellar mass
    is $M_{\rm star}=6.91 \times 10^{9} {\rm M}_{\odot}$.

\section{results}

\begin{figure}
	\includegraphics[width=1.1\linewidth]{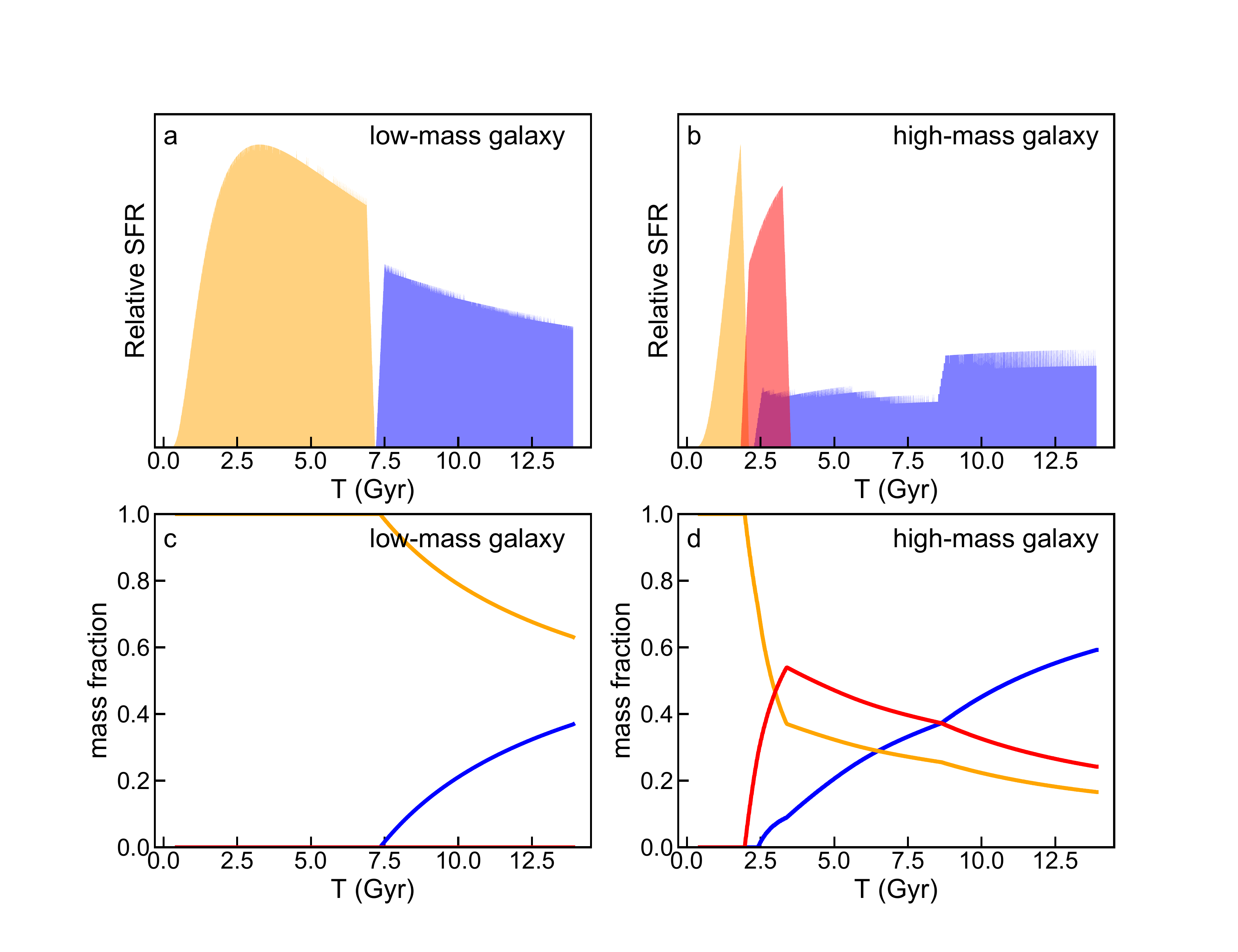}
    \caption{
	Panels a,b: Star formation rate (SFR) in the least massive and the most massive galaxies.
	Blue, orange, and red indicate SFR for thin disc, thick disc, and bulge, respectively. 
	SFR is averaged over 20 successive time steps (0.28 Gyr) to smooth out the short timescale fluctuations arising 
	in the model calculation, and given in arbitrary units. Panels c,d: Time variation of the mass fraction relative to 
	the total stellar mass of thin disc (blue), thick disc (orange), and bulge (red).}
    \label{Fig.2}
\end{figure}

Star formation history of the most and least massive galaxies in the model is shown in Fig.2.
As shown in Fig.2a, low-mass galaxies with $M_{\rm vir,0} < M_{\rm critical}$ first grow thick discs by the cold-mode accretion. 
After crossing the shock-heating border (i.e., $M_{\rm vir}=M_{\rm shock}$), the gas accretion is switched to the cooling flow (Fig.1), and thin discs are formed. 
Fig.2c shows that the mass fraction of the thick disc decreases monotonically with time. 
The massive galaxies with $M_{\rm vir,0} > M_{\rm critical}$
evolve in a more complicated way (Fig.2b). After the thick disc formation in Domain F, 
they enter the hybrid Domain H and develop bulges (Fig.1). Finally thin discs start to dominate when they leave this domain.
Galactic bulges in this scenario are not the first mass component to emerge  but 
thick discs predate bulges. Therefore, the bulge mass fraction is initially almost zero and then starts to increase as the bulge develops. 
In later phase (T > 7 Gyr), it decreases as the thin disc grows (Fig.2d).

\begin{figure}
	\includegraphics[width=1.1\linewidth]{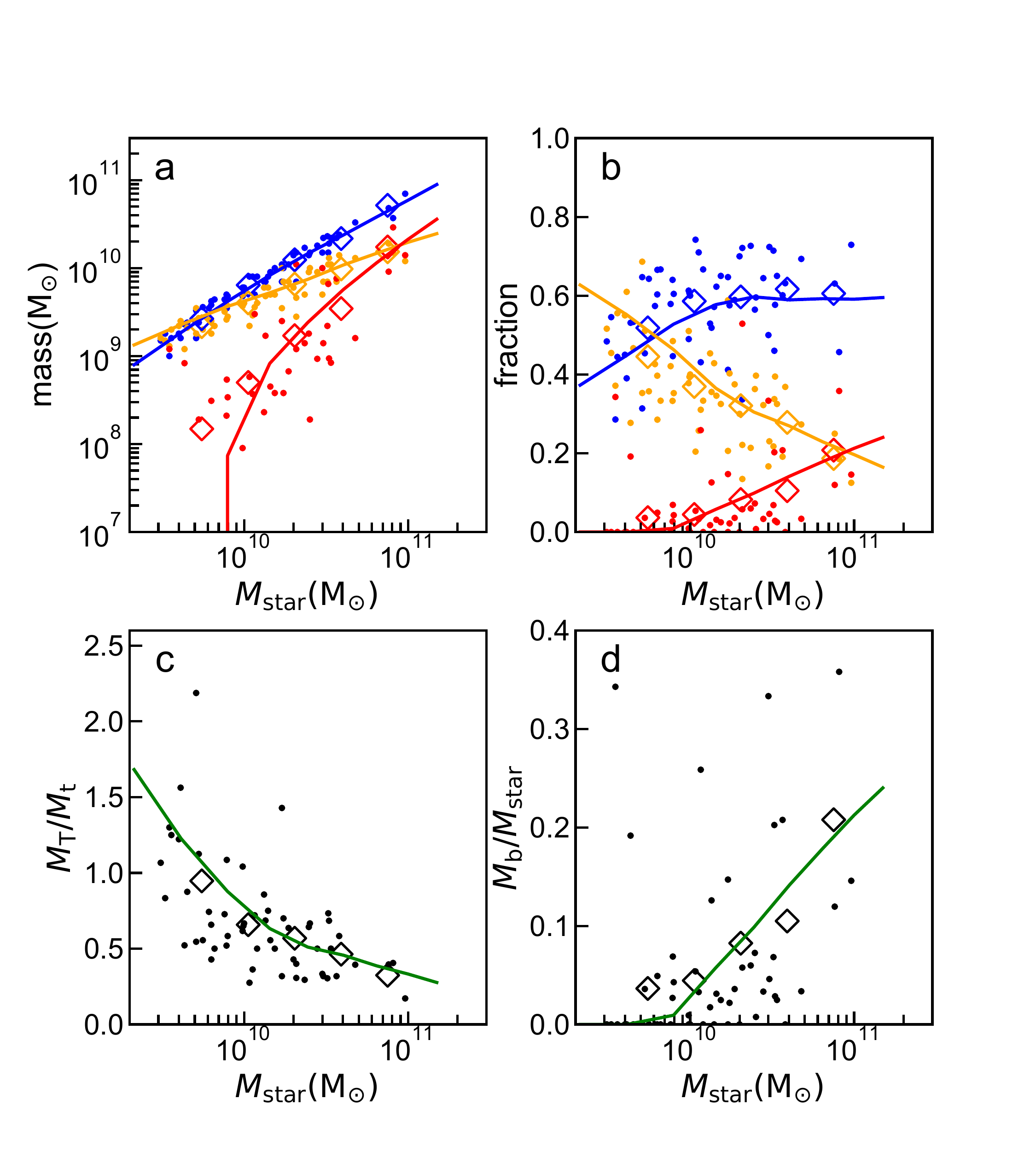}
    \caption{Masses of various structural components and their ratios plotted against the total stellar mass of the galaxy.
	The total stellar mass, $M_{\rm star}$, is the sum of the masses of thin disc ($M_{\rm t}$), thick disc ($M_{\rm T}$), 
	and bulge ($M_{\rm b}$). Solid lines show model results. The lowest and highest stellar masses
    in the model are $M_{\rm star}=2.13 \times 10^{9} {\rm M}_{\odot}$ and $1.50 \times 10^{11} {\rm M}_{\odot}$, respectively. 
    Observed values by \citet{co14} are indicated by dots, with diamonds showing 
	the mean in each mass-bin. Panel a: Mass of each component (thin disc: blue, thick disc: orange, bulge: red). 
	Panel b: The same as Panel a, but for the mass fraction of each component. \
    Panel c,d: Mass ratios involving the indicated components. }
    \label{Fig.3}
\end{figure}

Fig.3 summarizes the morphological observation performed by \citet{co14} which is the most extensive one at present.
Of the sample galaxies plotted in Fig.3, those with bulges are dominated by classical bulges and/or 
unresolved central mass components. We combined these components into ``bulge'' for comparison with the model.
A small number of galaxies with pseudo-bulges were excluded but including them does not 
make any significant changes. Boxy or peanut-shaped bulges were excluded because they are considered to be bar 
structures seen edge-on and therefore of disc origin. 

Fig.3a demonstrates that the model reproduces the absolute masses of various components as a function of 
the total galaxy stellar mass for the present-day disc galaxies reasonably well, although the observation shows a large scatter. 
Note that the model describes the mean behavior and does not take into account possible dispersions in the dark matter halo 
properties at the same virial mass. Both in the model and observations, the bulge mass shows the steepest increase as the total galaxy mass increases, 
while the thick disc mass increases mildly. The increase of the thin disc is the intermediate one.
Figs.3b,c, and d plot mass fractions and ratios against the galaxy mass.
The model agrees with the observation also regarding relative dominance of the different components.
Low-mass galaxies ($M_{\rm star} <10^{10} {\rm M}_{\odot}$) lack 
bulges or have only small bulges, with thick discs occupying comparable masses with thin discs  (Fig.3b). 
Galaxies with larger masses contain increasingly more massive bulge components but the contribution of thick discs decreases. 
The mass ratio of thick and thin discs decreases monotonically with the total stellar mass (Fig.3c), 
while the bulge-to-total mass ratio increases (Fig.3d). These results are consistent with the behavior of 
star formation rate in Fig.2 showing that massive galaxies experience more intense cooling flow accretion
in Domain G contributing to the thin disc growth in $T$ > 8.5 Gyr. They also experience  
an additional hybrid accretion in Domain H, which leads to the bulge formation by the cold gas flow around $T$ = 3 Gyr and 
the thin disc growth by the  cooling flow of the 
shocked gas during the period 2.5 Gyr < $T$ < 8.5 Gyr. 

\begin{figure}
	\includegraphics[width=1.\columnwidth]{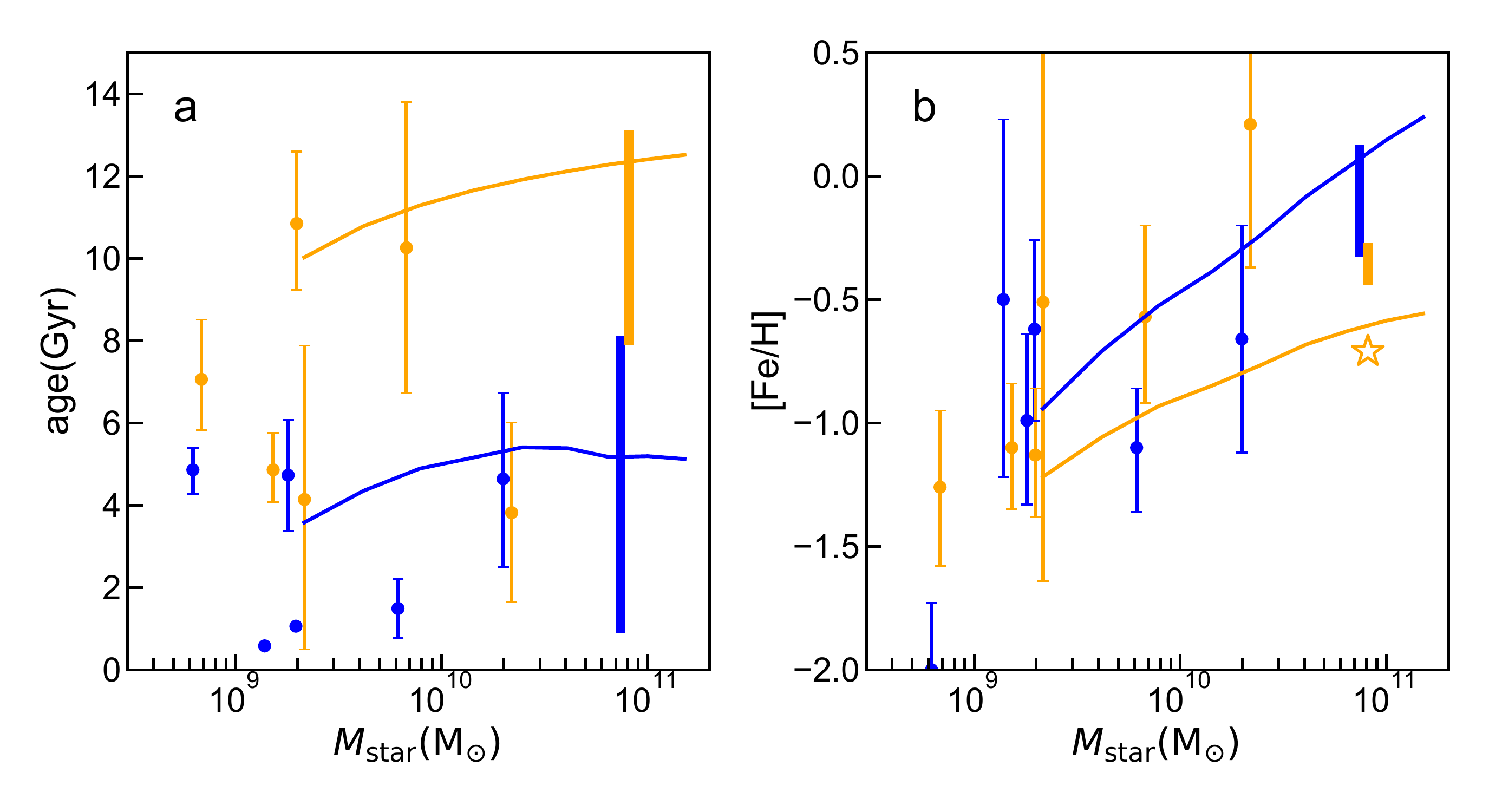}
    \caption{Age and iron abundance of the thick (orange) and thin (blue) discs plotted against the total stellar mass of 
	the galaxy. 
	Solid curves indicate the model. 
	Dots with error bars indicate the observations for external galaxies (thick disc values are slightly 
	displaced horizontally for clarity) by \citet{yo08}, with the galaxy circular velocity converted to the
	stellar mass using the Tully-Fisher relation by \citet{re11}.
	Panel a: The mean stellar age. The vertical thick segments indicate the age 
	range estimated for the solar neighbourhood data by \citet{ha13}. Panel b: The iron abundance ratio. The vertical thick segments 
	indicate the range for the solar neighbourhood data by \citet{ha15}. The star symbol indicates the abundance of 
    the solar neighborhood thick disc observed by \citet{re06}. }
    \label{Fig.4}
\end{figure}

The two kinds of discs are different also in age and chemical compositions as shown in Fig.4. 
Fig.4a shows that the model thick discs are old components with most stars formed more than 10 Gyr ago, 
whereas thin discs consist of much younger stars, in broad agreement with the observations for the solar 
neighbourhood and for other galaxies.
The chronological order of thick and thin disc formation is also reflected in their chemical properties. 
The model thin discs include more iron relative to hydrogen than the thick discs (Fig.4b) because they are formed 
from the interstellar gas polluted by iron produced and released by the supernovae
originating in early star formation which made the thick discs. 
Solar neighbourhood shows a similar trend, though the data for other galaxies show large scatters and no clear trend. 

We proposed here that the galactic thick discs were formed by the cold accretion. Other proposed mechanisms include 
the stirring-up of the existing thin discs by mergers with smaller galaxies \citep[e.g.][]{qu93} and the radial migration of 
kinematically hot stars from inner galactic regions to the solar neighbourhood \citep{mi13}. 
It seems, however, difficult to explain the distinct chemical bimodality of MW disc stars in these scenarios. 
Although several processes may contribute to the thick disc formation actually, the successful reproduction of 
structural variations (Fig.3) makes the cold-flow origin of thick discs a viable hypothesis, 
in combination with its ability to explain the age and chemical duality demonstrated in \citet{no18b}.
\citet{br09} investigated the accretion history of the different 
components of the numerically simulated disc galaxies and showed that the shock heated gas tends to 
end-up in thin discs, while the thick discs and bulges are predominantly formed from the gas accreted
in unheated states. Bulges also contain significant contributions from clump accretion.  
The present model is thus in broad agreement with the cosmological simulation by \citet{br09}, although the latter
does not discuss the change of accretion mode in relation to the growth of the host dark matter halo.

\section{Implication for high-redshift galaxies}

\begin{figure}
	\includegraphics[width=0.9\linewidth]{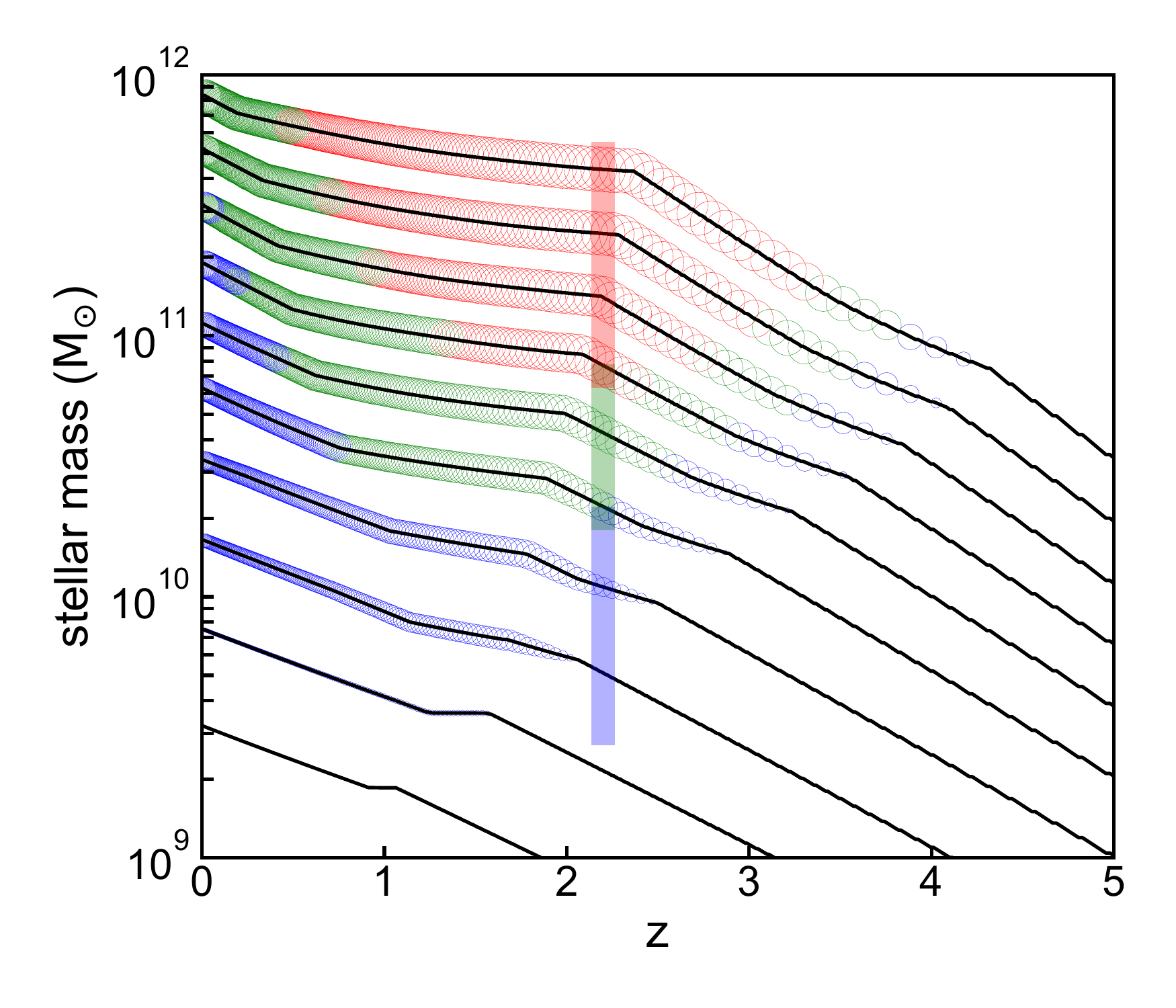}
    \caption{Evolution of the bulge-to-total stellar mass ratio for each model. Black lines show the evolution
	of the total stellar mass. Size of the circle is proportional to the bulge mass ratio,
     $f_{\rm b} \equiv M_{\rm b}/M_{\rm star}$, 
	 which is also colour-coded (blue: $f_{\rm b}$ < 0.3, green: 0.3 < $f_{\rm b}$ < 0.5,
	red: 0.5 < $f_{\rm b}$ < 0.8). 
    For this figure, the mass range of the halo is 
	$3.8 \times 10^{11}{\rm M}_{\odot}<M_{\rm vir,0}<2.5 \times 10^{13}{\rm M}_{\odot} $, 
	equaly spaced in $\log M_{\rm vir,0}$. Shaded segments indicate the observations by \citet{ta15} for z=2.2 galaxies 
	with the mean Sersic indices $n$ colour-coded (blue: $n=1$, green: $n=1.9$, red: $n=2.8$).}
    \label{Fig.5}
\end{figure}

Present study predicts a large structural variety of high-redshift galaxies. The particularly interesting 
 redshift range is 2<z<3, for which
we will witness three kinds of evolutionary status (Fig.1). Galaxies with the smallest masses will be in Domain F 
and making thick discs. Intermediate-mass galaxies, staying in Domain H, will be forming bulges in addition to the 
already-formed thick discs. Finally, the most massive galaxies will have already entered Domain G, creating thin discs. 
Present-day observations can reach such high redshifts, providing interesting new insights. 
For example, \citet{ta15} examined the star formation activity in galaxies at $z \sim 2$ and found that   
more massive galaxies show stronger quenching (i.e., cessation of star formation) in their central parts, 
hinting at the existence of matured bulges. This seems to agree partly with the above prediction. 
For more quantitative comparison, Fig.5 shows how the model galaxies build-up
bulge components as they grow in the stellar mass. 
\citet{ta15} measured the Sersic indices $n$ of the stellar density profile instead of the bulge mass
fraction, $f_{\rm b} \equiv M_{\rm b}/M_{\rm star}$. 
Based on the result by \citet{vi14}, we make a crude estimate that $n=1, 1.9$ and 2.8 correspond to $f_{\rm b}=0.2, 0.4$
and 0.6, respectively.
Once this conversion is made, the observation by \citet{ta15} agrees reasonably well with the model, with both 
showing the increasing dominance of the bulge for more massive galaxies.
In the present model, the bulge mass ratio attains a peak at $z \approx 2$, decreasing toward smaller redshifts due to
the growth of the thin disc and toward higher redshifts because thick discs predate the bulges.
Detection of this change both along the evolutionary tracks and at the fixed stellar masses by future observations
will further strengthen the scenario proposed in the present study.

\section{Conclusions}

We seek the origin of different structural components of disc galaxies in the different physical states 
of their raw material gas when it is distributed in the host halo.
We associate the shock-heated hot gas in massive halos with the thin disc components
and the cold unshocked gas in low mass halos with the thick discs. Further, we propose a new formation channel 
of the bulges from
the cold gas compressed by the surrounding shock-heated hot gas in massive galaxies  at high redshifts.
Calculation of the build-up process of each component for a wide galaxy mass range based on this 
proposed correspondence reproduces the
increasing dominance of the bulges and thick discs toward higher and lower galaxy masses, respectively. 
This scenario also explains the age and metallicity difference of thin and thick discs 
observed in the Milky Way and other disc galaxies.

\section*{Acknowledgements}

We acknowledge A. Faisst for providing data for the mass fraction of gas for star forming galaxies.
We thank anomymous referees for useful comments which helped us improve the paper.

%%%%%%%%%%%%%%%%%%%%%%%%%%%%%%%%%%%%%%%%%%%%%%%%%%



%%%%%%%%%%%%%%%%%%%%%%%%%%%%%%%%%%%%%%%%%%%%%%%%%%

%%%%%%%%%%%%%%%%% APPENDICES %%%%%%%%%%%%%%%%%%%%%

%%%%%%%%%%%%%%%%%%%%%%%%%%%%%%%%%%%%%%%%%%%%%%%%%%

% Don't change these lines
\bsp	% typesetting comment
\label{lastpage}
\end{document}

\label{firstpage}
\pagerange{\pageref{firstpage}--\pageref{lastpage}}
\maketitle

%%%%%%%%%%%%%%%%%%%%%%%%%%%%%%%%%%%%%%%%%%%%%%%%%%

%%%%%%%%%%%%%%%%% APPENDICES %%%%%%%%%%%%%%%%%%%%%

\section{Supplementary material: The disc galaxy evolution model}

In the present work, evolution of a disc galaxy is calculated in two steps.

In the first step, the gas accretion rate at each time step is calculated in the same manner as described 
in \citet{no18a}. We describe only the basic features of the original model and the modifications specific to 
the current version here. The growth of 
the halo virial mass with redshift $z$ is specified as $M_{\rm vir} (z)= M_{\rm vir,0}  e^{-2Kz/C}$ \citep{we02}, 
where $M_{\rm vir,0}$ is the present virial mass, $K$ is a constant 3.7, and the halo concentration parameter $C$ 
at the present epoch is given by 
\begin{equation}
	{\log} C=0.971-0.094  {\log}  (M_{\rm vir,0}  / 10^{12}  h^{-1} {\rm M}_{\odot})
\end{equation}
following \citet{bu01}.
The gas gathered by the halo is assumed to have the same Navarro-Frenk-White density profile as the halo, 
with the adopted cosmic baryon fraction of 0.17. 
The gas accretion rate in the shock-heating regime  (Domain G) is calculated 
by assuming that the gas is heated to the virial temperature when entering the host halo and accretes to the disc 
plane with the shorter of the radiative cooling or the freefall times. 
The cooling time is calculated at the virial radius when the gas enters the halo
using the collisional ionization equilibrium cooling function of \citet{su93}, whereas the 
free-fall time is calculated from the mean total density inside the virial radius.
The accretion rate in the cold-flow regime (Domain F) is calculated assuming that the gas accretes 
with the freefall time. In Domain H, we assume that half the gas 
during this period accretes as cold flows and another half as cooling flows. 
Instead of resolving the galaxy into 
 a series of concentric annuli as done in the previous studies \citep{no18a,no18b}, we neglect its spatial structure here and therefore do not consider the angular momentum distribution of the gas. 
 The calculated gas accretion rate thus gives the total accretion rate at each time.

The radiative cooling time is sensitive to the metallicity of the halo gas, which 
is not well constrained observationally and is likely to depend on the redshift and radial position.
Observations for intergalactic medium hint at an outer halo metallicity $Z \sim 0.003Z_{\odot}$ with no significant 
evolution \citep{sc03}. On the other hand, the inner halo metallicity probed by the damped Lyman systems seems 
to evolve from $Z \sim 0.01Z_{\odot}$ at $z=4$ to $Z \sim 0.1Z_{\odot}$ for  $z<1$ \citep{pr03}. For the present one-zone
model, we adopted the metallicity of the halo gas of 
$0.01Z_{\odot}$ as a crude mean and used this constant value 
in calculating the cooling time of the virialized halo gas. 
This is smaller than $0.1Z_{\odot}$ adopted in \citet{no18a} and   
leads to a lower $M_{\rm shock}$. Guided by the result for 
0.1 $R_{\rm vir}$ presented in Fig.2 of \citet{de06}, 
we take $M_{\rm shock} = 1.5 \times 10^{11} {\rm M}_{\odot}$ independently of the redshift. 
$M_{\rm stream}$ also carries a large uncertainty. The value analytically derived  by \citet{de06} using 
the present metallicity $Z_0=0.1$ is about 1.7 dex larger than the one obtained in the cosmological 
simulation by \citet{oc08} at the same redshift. We take intermediate values between these two as indicated in Fig.1, 
which are closer to Ocvirk's values. 

In the second step, the stellar mass growth of the thin disc, the thick disc, and the bulge is calculated, 
assuming that the star formation rate is proportional to the accretion rate.
The total mass of the stars formed is 
adjusted to the observed stellar mass at present using the relationship between the halo mass and stellar mass 
for late-type galaxies given by eq(3) in \citet{du10}. 
In this process, the proportinality coefficient
is held common to all the galactic components in each galaxy but is allowed to be different for different models
to satisfy the observed $M_{\rm vir}-M_{\rm star}$ relation. 
Note that the accretion regimes in this study are completely defined by two parameters, $M_{\rm shock}$ and $M_{\rm stream}$ 
(which are held fixed),
so that other parameters  used in equations (18) and (19) of \citet{no18a} do not appear here.

Not all the gas accreted is used to form stars in real galaxies. 
Indeed, the total gas mass that accretes during the entire time in the model is larger than 
the stellar mass observed. In actual galaxies, a significant portion is considered to escape by the stellar 
(supernova) feedback or the action of active galactic nuclei. Detailed mechanisms and efficiency of gas removal 
in these feedback processes are much debated. Instead of tackling this problem, we choose to study how 
each mass component of the galaxy is formed under the constraint that the total mass of produced stars is 
consistent with the observed values. In other words, the present treatment includes possible effects of various
feedback processes in an implicit form.
  
In order to follow the chemical element abundances in the interstellar gas and stars, we must specify the mass of 
cold interstellar gas (i.e., molecular hydrogen) at each moment and introduce the metal enrichment by supernovae. 
Mass of the cold gas is assumed to follow the observationally inferred values obtained by \citet{fa17} for a redshift range 0<z<2. 
Specifically, the cold gas fraction for the stellar mass $M_{\rm star}= 10^{10.25},10^{10.75}$, and $10^{11.4} {\rm M}_{\odot}$ is set 
to 0.11, 0.09, and 0.05 at z=0, and 0.42, 0.32 and 0.22 at z=2, respectively. 
The gas fraction at each time step was interpolated (or extrapolated if necessary) from these reference values 
using the current redshift and stellar mass of the model. Lower and upper cutoff of the gas fraction is imposed 
at 0.05 and 0.9, respectively. 

Chemical enrichment due to Type II and Ia supernova (SN) is treated in the same way as in the previous study 
except slight modifications of stellar yields and Type II SN lifetime. Namely, $\alpha$-elements of 3.18 ${\rm M}_{\odot}$ and 
iron of 0.094 ${\rm M}_{\odot}$ are added to the interstellar gas for one Type II SN. One Type Ia SN is assumed to 
return 1.38 ${\rm M}_{\odot}$ of material to the interstellar gas, of which $\alpha$-elements and iron occupy 0.438 ${\rm M}_{\odot}$  
and 0.74 ${\rm M}_{\odot}$ respectively \citep{fe02}. Type II SNe are assumed to explode not instantaneously as done in 
the previous works, but $10^7$ years after the birth of progenitor stars. 



%%%%%%%%%%%%%%%%%%%%%%%%%%%%%%%%%%%%%%%%%%%%%%%%%%

% Don't change these lines
\bsp	% typesetting comment
\label{lastpage}